%% file: manuscript.tex
\documentclass[%
superscriptaddress,
 amsmath,amssymb,
 aps,
 prstab, twocolumn,
longbibliography,
]{revtex4-1}

\usepackage{tabularx}
\usepackage{textcomp}
\usepackage{graphicx}%
\usepackage{framed}
\usepackage[dvipsnames]{xcolor}
\usepackage{appendix}
\usepackage{amsmath}
\usepackage{amssymb}
\usepackage{dcolumn}
\usepackage{graphicx}
\usepackage{longtable}
\usepackage{bm}
\usepackage{color}
\usepackage{epsfig}
\usepackage[colorlinks, citecolor=blue, urlcolor=black, bookmarks=false, linkcolor=blue, hypertexnames=true]{hyperref}

\begin{document}

\title{Foray into the topology of poly-bi-[8]-annulenylene}
\affiliation{Department of Physics, Purdue University, West Lafayette, Indiana 47907, USA}
\affiliation{Department of Chemistry, Purdue University, West Lafayette, Indiana 47907, USA}
\affiliation{Purdue Quantum Science and Engineering Institute, Purdue University, West Lafayette, Indiana 47907, USA}

\author{Varadharajan Muruganandam}
\affiliation{Department of Physics, Purdue University, West Lafayette, Indiana 47907, USA}
\affiliation{Purdue Quantum Science and Engineering Institute, Purdue University, West Lafayette, Indiana 47907, USA}

\author{Manas Sajjan}
\affiliation{Department of Chemistry, Purdue University, West Lafayette, Indiana 47907, USA}
\affiliation{Purdue Quantum Science and Engineering Institute, Purdue University, West Lafayette, Indiana 47907, USA}

\author{Sabre Kais}
\email{kais@purdue.edu}
\affiliation{Department of Physics, Purdue University, West Lafayette, Indiana 47907, USA}
\affiliation{Department of Chemistry, Purdue University, West Lafayette, Indiana 47907, USA}
\affiliation{Purdue Quantum Science and Engineering Institute, Purdue University, West Lafayette, Indiana 47907, USA}

\begin{abstract}

Analyzing phase transitions using the inherent geometrical attributes of a system has garnered enormous interest over the past few decades. The usual candidate often used for investigation is graphene- the most celebrated material among the family of tri co-ordinated graphed lattices. We show in this report that other inhabitants of the family demonstrate equally admirable structural and functional properties that at its core are controlled by their topology. Two interesting members of the family are Cylooctatrene(COT) and COT-based polymer: poly-bi-[8]-annulenylene both in one and two dimensions that have been investigated by polymer chemists over a period of 50 years for its possible application in batteries exploiting its conducting properties.  A single COT unit is demonstrated herein to exhibit topological solitons at sites of a broken bond similar to an open one-dimensional Su-Schrieffer-Heeger (SSH) chain. We observe that Poly-bi-[8]-annulenylene in 1D mimics two coupled SSH chains in the weak coupling limit thereby showing the presence of topological edge modes. In the strong coupling limit, we investigate the different parameter values of our system for which we observe zero energy modes. Further, the application of an external magnetic field and its effects on the band-flattening of the energy bands has also been studied. In 2D, poly-bi-[8]-annulenylene forms a square-octagon lattice which upon breaking time-reversal symmetry goes into a topological phase forming noise-resilient edge modes. We hope our analysis would pave the way for synthesizing such topological materials and exploiting their properties for promising applications in optoelectronics, photovoltaics, and renewable energy sources.   


\end{abstract}
\maketitle
\section*{Introduction}

From the discovery of the quantum hall effect in the 1980s, \cite{klitzing1982, Tsui1982}, the perception of phases in condensed matter physics underwent a foundational metamorphosis. Phase transition in such systems formerly studied through the lens of Landau's theory of symmetry breaking \cite{ter2013collected, hoffmann2012ginzburg}, were subsequently analyzed using abstruse yet mathematically elegant characterization of the inherent geometrical attributes of the system thereby initiating a robust bridge to topology\cite{bredon2013topology}. 
Such inter-connections have positively impacted many other domains of physics including atomic physics and quantum optics\cite{diehl2011topology, PhysRevLett.119.023603,barik2018topological,amo2018quantum,barik2020chiral}, bio-informatics \cite{adams2020knot,mishra2012knot,sulkowska2009dodging}, quantum field theory \cite{witten1988topological}, high-energy physics \cite{donnelly2021entanglement,cole2019topological} and astronomy \cite{wei2019intrinsic,wei2020topological} even though condensed matter physics indisputably continues to be the most ardent and persistent beneficiary. A quintessential example in the latter domain which has arrested enormous attention over the past several decades are the family of organic polymers like polyacetylene\cite{shirakawa} which possesses albeit simple yet rich topological features in 1D\cite{Meier_2016} rooted in the Su-Schrieffer-Heeger (SSH) model \cite{ssh, Nobelprize2000}. Discovery of such polymers has revolutionized diverse applications like molecular electronics \cite{lakshmi2008molecular,jean2006inelastic, ness2005vibrational}, light-emitting diodes (LEDs) \cite{PhysRevLett.86.3602,meng2009recombination}, rechargeable batteries \cite{PhysRevB.61.1096,goto2008preparation,heeger2001semiconducting} to name a few, owing to their fascinating conducting properties usually accredited to the implicit topology and lattice geometry.

\begin{figure*}[t]
\begin{framed}
\centerline{
\includegraphics[width = 5.86 in]{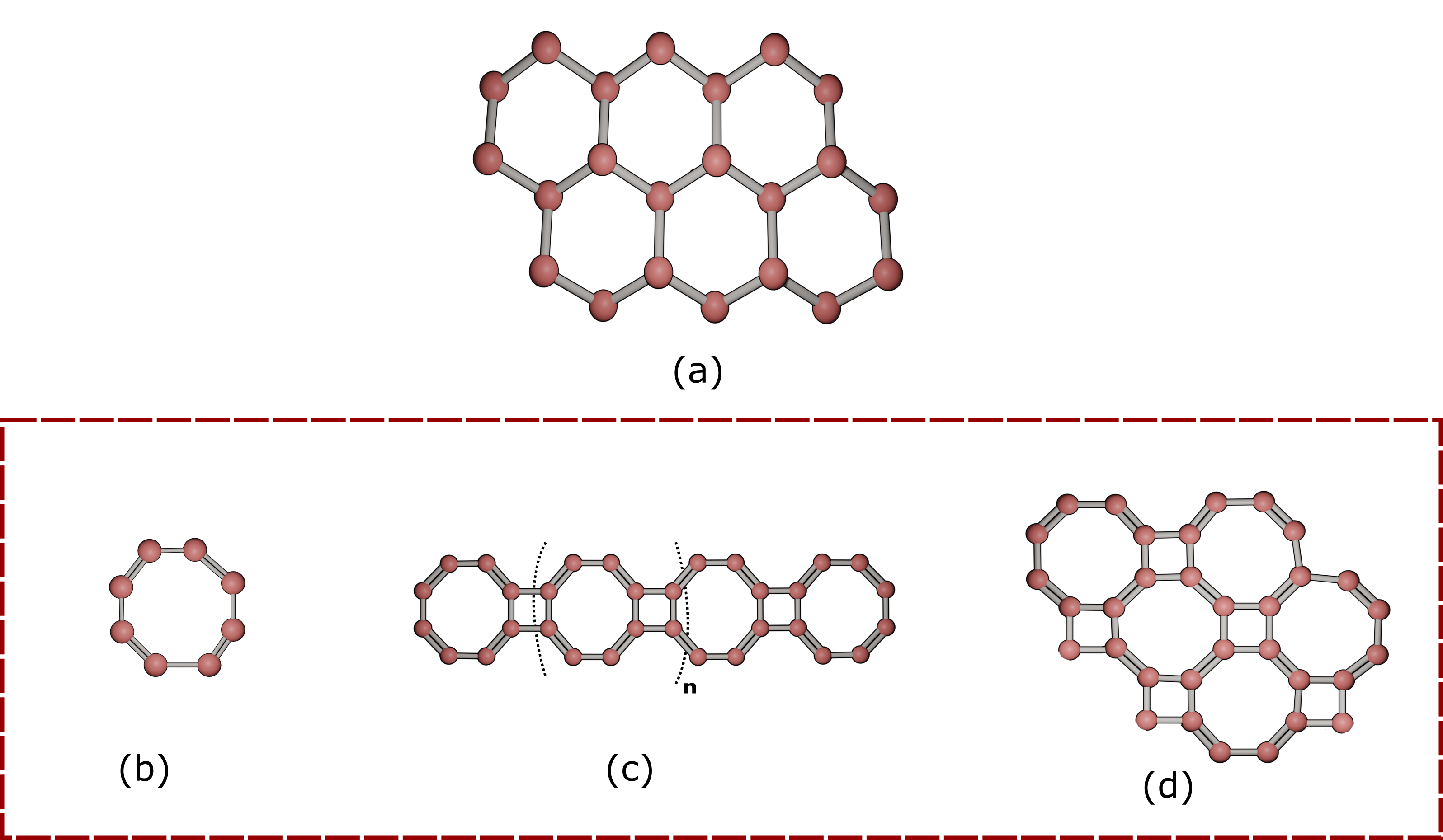}}
\caption{
(a) Graphene
(b) cycloctatetraene(COT) unit
(c) repeating Poly-bi-[8]-annulenylene in 1D
(d) repeating Poly-bi-[8]-annulenylene in 2D.
(red box: Goldilocks zone of COT and COT based polymers)
}\label{fig:fig1}
\end{framed}
\end{figure*}

The natural extension of the aforesaid paradigm to 2D began with the idea of Haldane \cite{haldane} which introduces a complex second nearest-neighbor hopping amplitude in Graphene, which is inarguably the most widely known honeycomb lattice belonging to the larger umbrella of trivalent graphed lattices (i.e. lattice geometries with co-ordination number equal to 3)  as shown in Fig. \ref{fig:fig1}. The by-product of such an endeavor is the decimation of the time-reversal symmetry(TRS) of the system thereby culminating in a natural emergence of a topological phase that is experimentally realizable.\cite{Jotzu_2014}. Extension of the paradigm to structural chemistry has been the harbinger of a plethora of unforeseen opportunities that has duly engendered interest \cite{chakraborty2021two,jiang2021exotic,gao2020design,jiang2020topological}. Most notably with prodigious improvements in synthetic capabilities of metal-organic and covalent-organic frameworks (MOFs/COFs) \cite{poly1,poly2,poly3,poly4,FBORGANIC,FBORGANIC2,Su2018}, the dream of artificially designing such polymeric substrates with tunable topological features is no longer distant.
Inspired by such developments, in this work, we strive to venture beyond graphene into other members of the family of trivalent graphs which despite having the potential for offering tantalizing prospects have been severely under-utilized in literature. We focus on the Goldilocks zone of such polymers (marked in red in Fig. \ref{fig:fig1}) which have either been directly synthesized and shown to be excellent conductors as highlighted in a patent\cite{patent} and paper\cite{COTZhao2018SuperposedRC} or offers an easy possibility of being naturally synthesizable or artificially designed through a network of superconducting coplanar waveguides \cite{alicia1}. Such lattices share similar structural cohomology with the hexagonal lattice of graphene \cite{archimeded1,chavey1989tilings} and as we shall unravel also inherit some exotic topological features even within the framework of tight-binding(TB) approximations which form the basis to interpret all their functional attributes.

The article is structured as follows. First, we consider a single Cycloctatetraene(COT) unit which forms the basic building block for Poly-bi-[8]-annulenylene networks, an object of primary investigation in this work. We show
how a COT unit forms topological solitons at two ends by envisioning it as a simple 1D SSH chain. Then We go on to study the bandstructure of Poly-bi-[8]-annulenylene in 1D and unravel its inherent topological properties both in the strong and weak coupling regime. In the weak regime, we consider Poly-bi-[8]-annulenylene as two weakly coupled 1D SSH chains. Following that, we further analyze the effects of band-flattening of all the energy bands of 1D Poly-bi-[8]annulenylene in the presence of an external magnetic field. Such band-flattening can lead to localized electronic states with exotic correlated behavior. We also study the 2D extension of this lattice geometry and calculate its band structure both with and without breaking TRS in order to distinguish topologically trivial and non-trivial phases. All analysis is conducted for both periodic(PBC) and open-boundary conditions (OBC) and the possibility of engineering a spin network for realizing the 2D analog is explicitly discussed.

\section*{RESULTS \& DISCUSSION}

\subsection*{Cycloctatetraene(COT) as a SSH chain}\label{sec1}
\begin{figure}[h]
\centerline{
\includegraphics[width = 3 in]{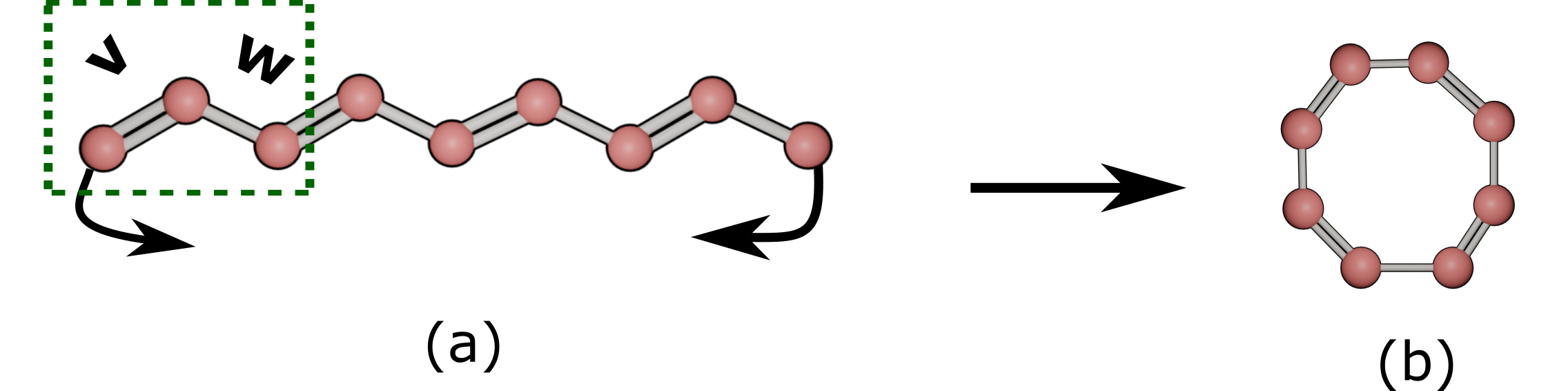}}
\caption{
(a) 1D SSH chain (a unit cell encompassed in the green box)
(b) COT as a closed SSH chain
}\label{fig:figssh1}
\end{figure}

\begin{figure}[t]
\centerline{
\includegraphics[width = 3.3 in]{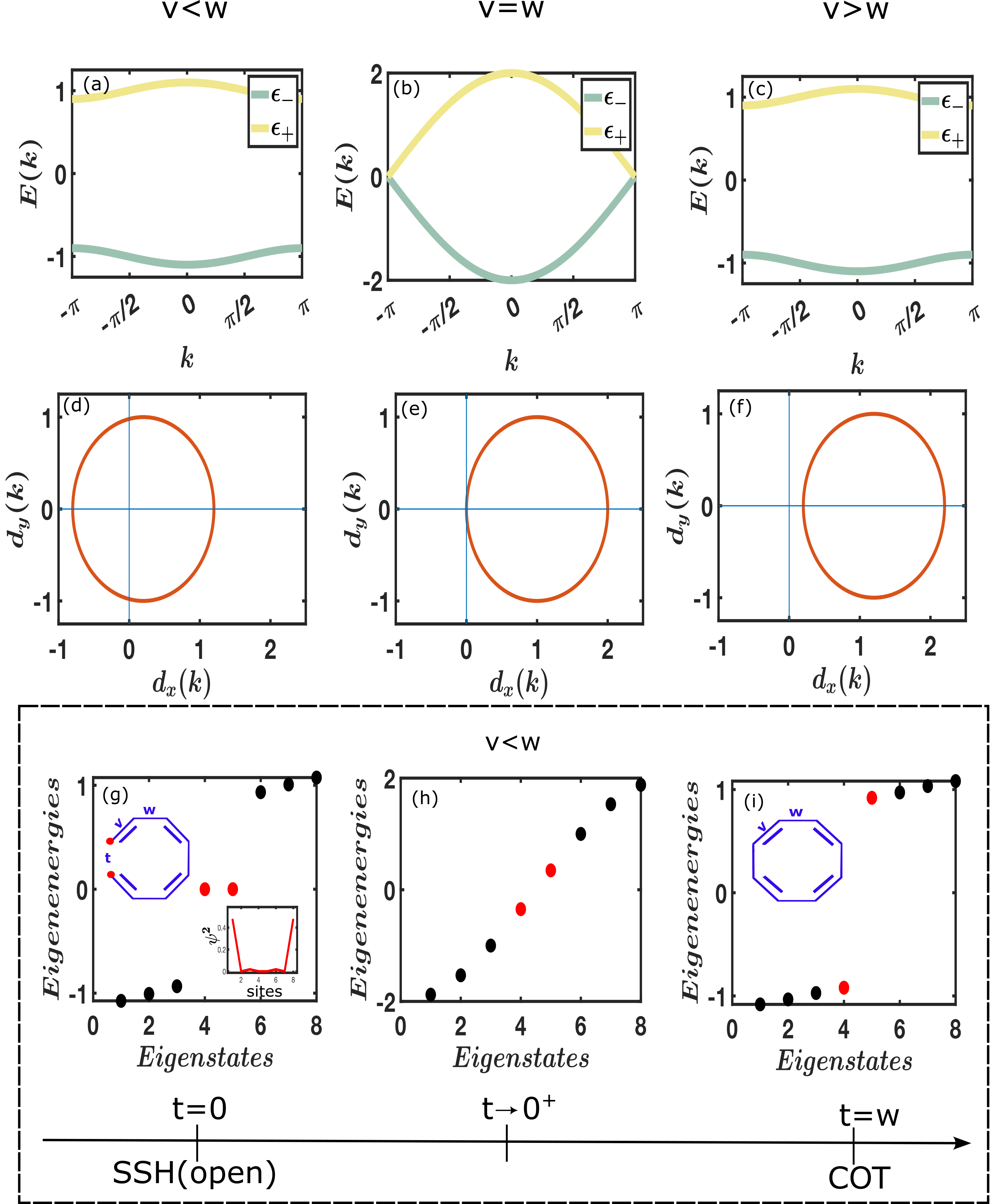}}
\caption{
In Fig\ref{fig:ssh_main}(a-c) we explicate the band structure of the model described in Eq. \ref{Ham_ssh_1D} and Eq. \ref{d_vec_1D_ssh}. 
(a) The spectrum ($E(k)\:\: \rm{vs} \:\: k$) under periodic boundary conditions (PBC) in the topological phase, (b) Similar to (a) but at the critical point showing the closure of the bulk gap (c) Similar to (a) but in the trivial phase.
(d) The locus of the vector $\vec{d_k}$ (see Eq. \ref{d_vec_1D_ssh}) in the topological phase with the origin enclosed, (e) at the critical point where the curve goes through the origin, (f) in the trivial phase with the origin not enclosed.The energy spectrum under open boundary conditions (OBC) (g) showing the presence of edge modes in the topological phase (h) in the bulk-conducting phase (i) band-insulating (topologically trivial) phase. The inset in (g) shows the electronic density distribution corresponding to the edge modes being localized at the edges of the chain.
}\label{fig:ssh_main}
\end{figure}

To explore the topology of COT, we begin by considering a single COT unit as a closed and periodic SSH chain. The well-known SSH chain offers a paradigmatic example of supporting a one-dimensional topological insulating phase. The Hamiltonian of the model is as follows\cite{ssh}, 
\begin{align}
H_{SSH}=&-v\sum_{i=1}^Nc_{2i-1}^\dagger c_{2i}-w\sum_{i=1}^Nc_{2i}^\dagger c_{2i+1}+h.c \label{Ham_ssh_1D}
\end{align}
where $v,\,w$ are the alternating hopping strengths and $N$ denotes the number of unit cells with a single unit cell marked in green as shown in Fig. \ref{fig:figssh1}(a) $c^\dagger,\,c$ denote the creation and annihilation operators describing an electron hopping in a lattice between sites designated as $i$ and $h.c$ denotes hermitian conjugate. 
The Hamiltonian can be diagonalized through a fourier transformation\cite{Bloch1929berDQ,ashcroft1976solid},
\begin{equation}
H_{SSH}=-\sum_{k}\psi_k^\dagger H_k\psi_k=-\sum_{k}\psi_k^\dagger (\vec{d_k}.\vec{\sigma})\psi_k
\end{equation}
where $k*a_0 \in [-\pi, \pi]$ denotes a discrete set of points over the Brillouin zone in reciprocal space with $a_0$
being the real space lattice separation. ($\psi_k^\dagger,\psi_k)$ and $H_k$ denote Bloch vectors and Hamiltonian in the reciprocal space respectively. The $\vec{d_k}$ and the energy of the bands are expressed as,
\begin{equation}
\vec{d_k}=(v+wcos(kx),wsin(kx),0)
\end{equation}
\begin{equation}
\epsilon_{\pm}=\sqrt{d_x^2+d_y^2}=\pm\sqrt{v^2+w^2+2vw\,cos(k)} \label{d_vec_1D_ssh}
\end{equation}
  As shown in Fig. \ref{fig:ssh_main}, We have the trivial insulator phase for $v > w$, the metallic phase for $v = w$, and the topological insulator phase for $v < w$. The topological phase is further characterized by the non-zero winding number\cite{Asboth}(Methods for calculation of winding number) and the appearance of edge states. 

\subsection*{Poly-bi-[8]-annulenylene(PO[8]A) in 1D} \label{sec2}
\begin{figure}[h]
\centerline{
\includegraphics[width = 2.5 in]{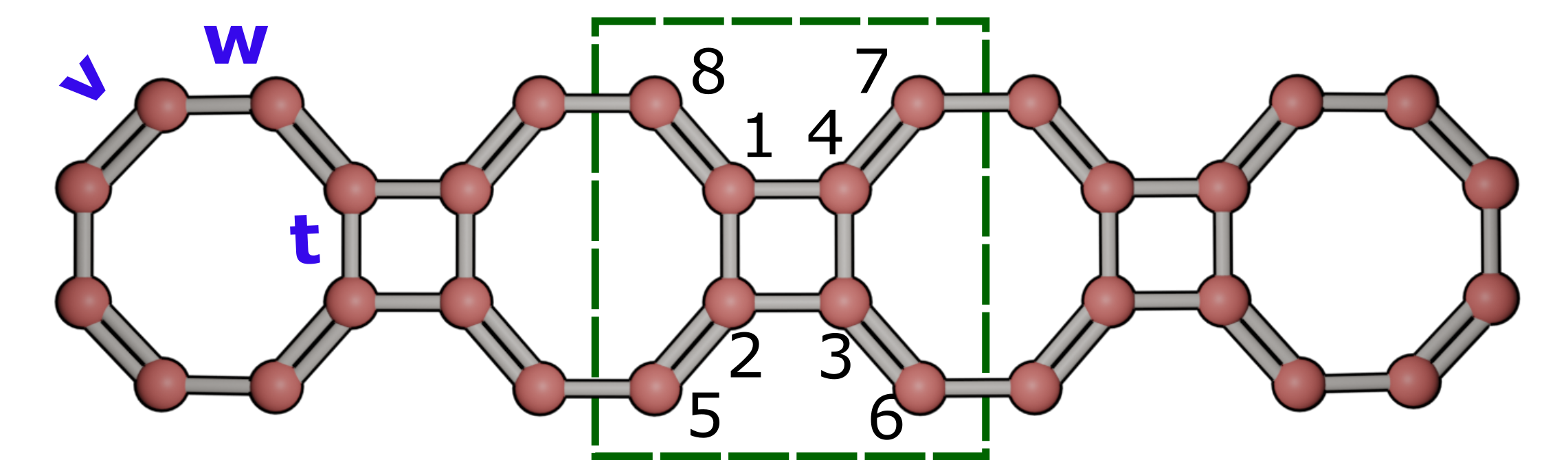}}
\caption{
Poly-bi-[8]-annulenylene(PO[8]A) moeity with the chosen unit cell shown inside the rectangular box. $v,\,w$ and $t$ shown are the respective hopping parameters. 
}\label{fig:poly}
\end{figure}

\begin{figure}[h]
\centerline{
\includegraphics[width = 3.3 in]{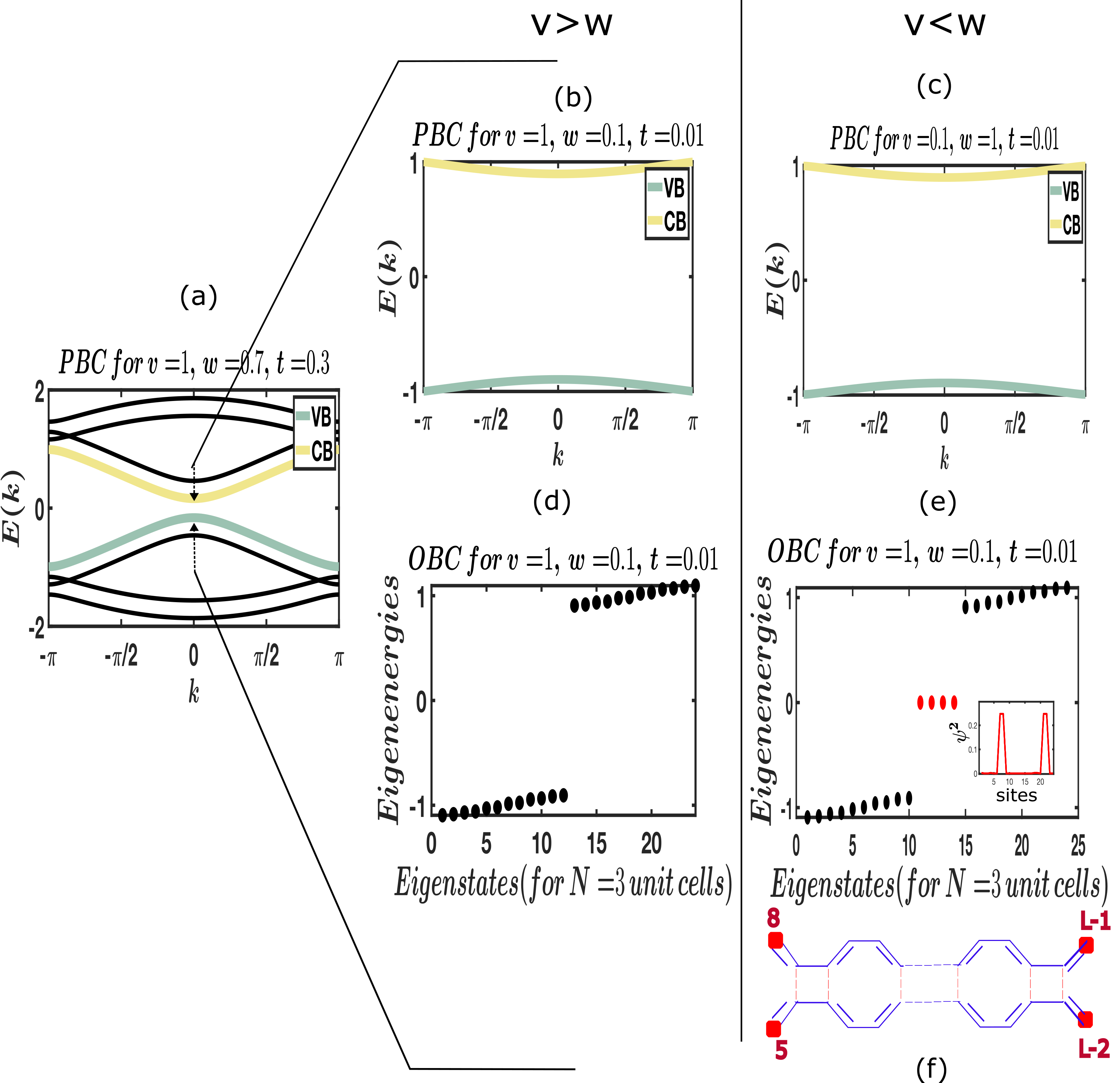}}
\caption{
(a) bandstructure of Poly-bi-[8]-annulenylene(PO[8]A) with $v=w=1,t=0.1$ highlighting the behavior of two weakly-coupled SSH chains under PBC.
(b) The spectrum as in (a) but computed in the trivial phase ($v>w$), (c) The spectrum as in (a) but computed in the topological phase ($v<w$). All plots are generated in the weakly coupled regime as defined by $t<<w$ and $t<<v$.
(d)The energy spectrum under OBC showing the absence of edge modes at zero energy in the trivial phase (e) Similar spectrum as in (d) but showing the presence of edge modes at zero energy in the topological phase. The inset shows electronic density distribution of the edge modes being localized at the four edges of a finite Poly-bi-[8]-annulenylene. 
(f)four edges of a finite Poly-bi-[8]-annulenylene(PO[8]A). 
}\label{fig:ssh2}
\end{figure}

For Poly-bi-[8]-annulenylene, the unit cell marked in green in Fig. \ref{fig:poly}, has eight sites. The Hamiltonian of the model is as follows,
\begin{align}
    H_{PO[8]A}=&\sum_{r=1}^N[-v(c_{r,1}^\dagger c_{r,8}+c_{r,4}^\dagger c_{r,7} \nonumber \\
    &+c_{r,2}^\dagger c_{r,5})-w(c_{r,1}^\dagger c_{r,4}+c_{r,2}^\dagger c_{r,3}) \nonumber \\
    &-t(c_{r,1}^\dagger c_{r,2}+c_{r,4}^\dagger c_{r,3})]+h.c \label{eq:poly_model}
\end{align}
where $v$, $w$, and $t$ are the respective hopping strengths between different lattice sites in a unit cell as shown in Fig. \ref{fig:poly}(a). The subscript $r,i$ contains $i=1..8$ going over all the eight sites in a unit cell and $r$ going over $N$ number of unit cells. Following a fourier transformation, We solve the k-space hamiltonian with the following form, 
\begin{align}
H_{PO[8]A}=-\sum_{k}\psi_k^\dagger [H(k)_{8\times8}]\psi_k
\end{align}
The corresponding band structure has eight bands as shown in Fig. \ref{fig:ssh2}(a). Among the eight bands, we focus on the valence and conduction band (marked as VB and CB) around the fermi-level(i.e. $E(k)=0$ level). We focus on two cases. \newline
\textit{Weak Coupling:} When the coupling strength $t<<v$ and $t<<w$, one can envision the lattice of Poly-bi-[8]-annulenylene as two weakly coupled SSH chains\cite{ssh2chains}- \cite{smitha1} as in \ref{fig:poly}(b) with parameter $t$ playing the role of linking between two chains. 

\begin{figure}[h]
\centerline{
\includegraphics[width = 3.3 in]{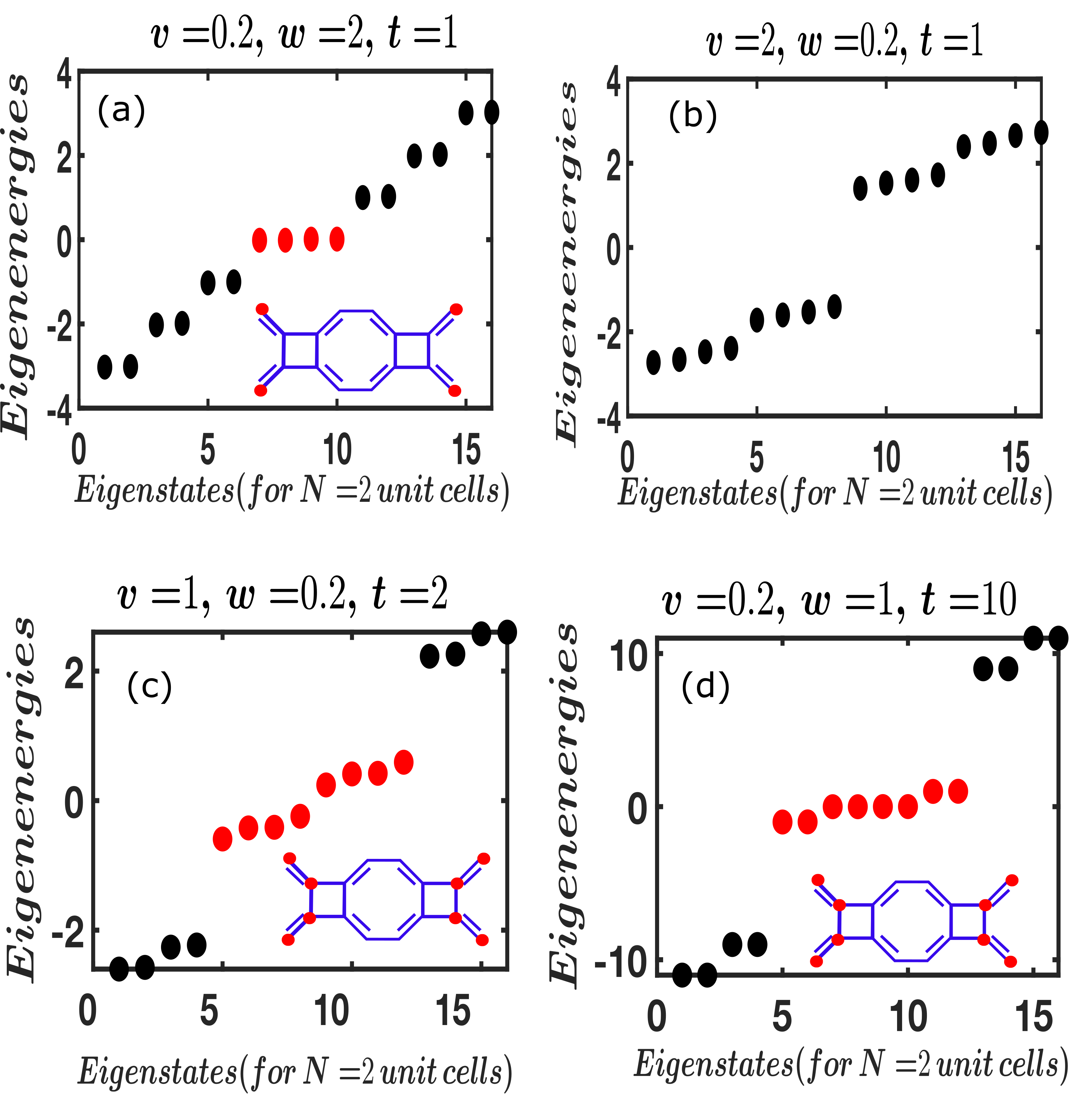}}
\caption{
(a)-(d)The energy spectrum with open boundary conditions showing the absence and presence  of zero modes in different parameter regimes.
The insets in (a)-(c) showing eight edges of a finite Poly-bi-[8]-annulenylene for N=2 unit cells  
}\label{fig:ssh2open}
\end{figure}

In Fig. \ref{fig:ssh2}, we show the band structure of the model defined in Eq. \ref{eq:poly_model} focusing exclusively on the energy bands close to the fermi-level in both the trivial and topological phase under PBC (Fig. \ref{fig:ssh2}(b-c)) and OBC (Fig. \ref{fig:ssh2}(d-e)). In the topological phase under OBC we expect four edge states corresponding to four edges of two weakly coupled SSH chains. This is further corroborated through numerical evidence in Fig. \ref{fig:ssh2}(e). Further, the resilience of these modes against noise has been discussed in Supplementary. This observation can be generalized to $N$ weakly coupled SSH chains wherein $2N$ edge modes would be encountered. 
\textit{Strong Coupling:} When the coupling is strong, the two stacked chains form dimers in several ways. Several such cases have been shown in Fig. \ref{fig:ssh2open}. All such cases are constructed under OBC and show the relative displacement in the energy spectrum of the eight edge states (see Fig. \ref{fig:ssh2open}(b),(d)) under different parameter choices. In general for $N$ such strongly coupled chains, one can prepare a maximum of $8N$ such edge states in the topological phase.


\subsection*{Magnetix flux through 1D PO[8]A}\label{bflux}
In this section, We study how the flatness of the Bloch bands could be tuned by magnetic fluxes through the square and octagon plaquettes of Poly-bi-annuleneylene. In the presence of an external magnetic field, the wave function of a charged particle going around a closed loop acquires an Aharonov-Bohm phase shift proportional to the magnetic flux through the enclosed loop\cite{bohm}. In the tight-binding(TB) language, this phase gets reflected in the electron hopping amplitudes modified with an acquired extra phase factor given by the Pierls substitution\cite{peierls},
\begin{equation}
    T_{ij}\rightarrow T_{ij}e^{\pm i\phi}
\end{equation}
Where $T_{ij}$ is the hopping amplitude between sites $i$ and $j$. For convention, We take positive flux in the clockwise direction and vice-versa.
\begin{figure}[h]
\centerline{
\includegraphics[width = 1.6 in]{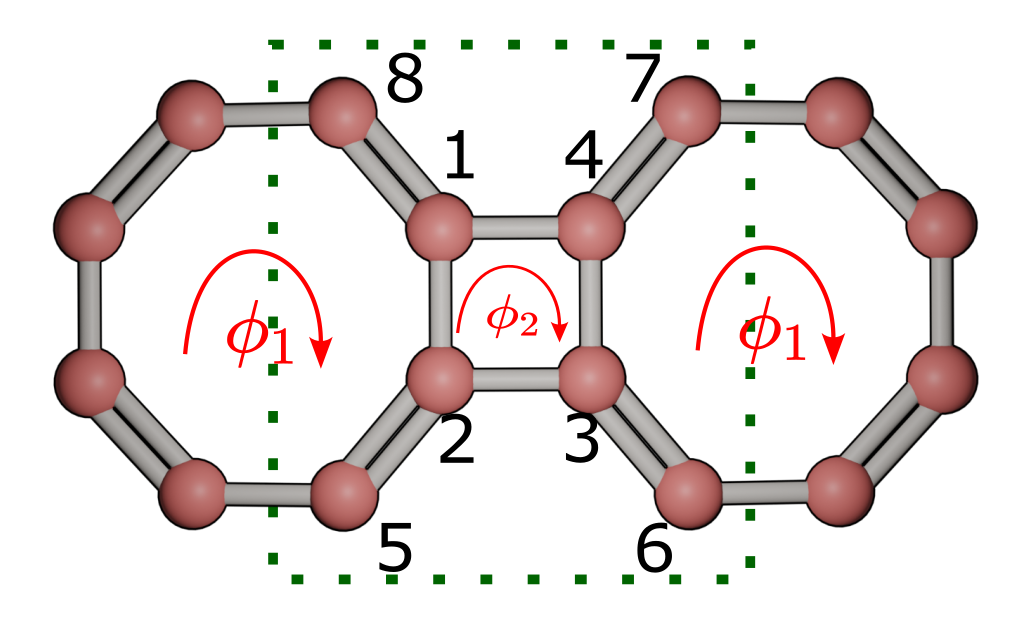}}
\caption{
Fluxes $\phi_2$ through the square and $\phi_1$ through Octagon plaquettes along the chosen clockwise direction. green box: unit cell of Poly-bi-[8]-annulenylene
}\label{fig:flux}
\end{figure}

\begin{figure}[h]
\centerline{
\includegraphics[width = 3 in]{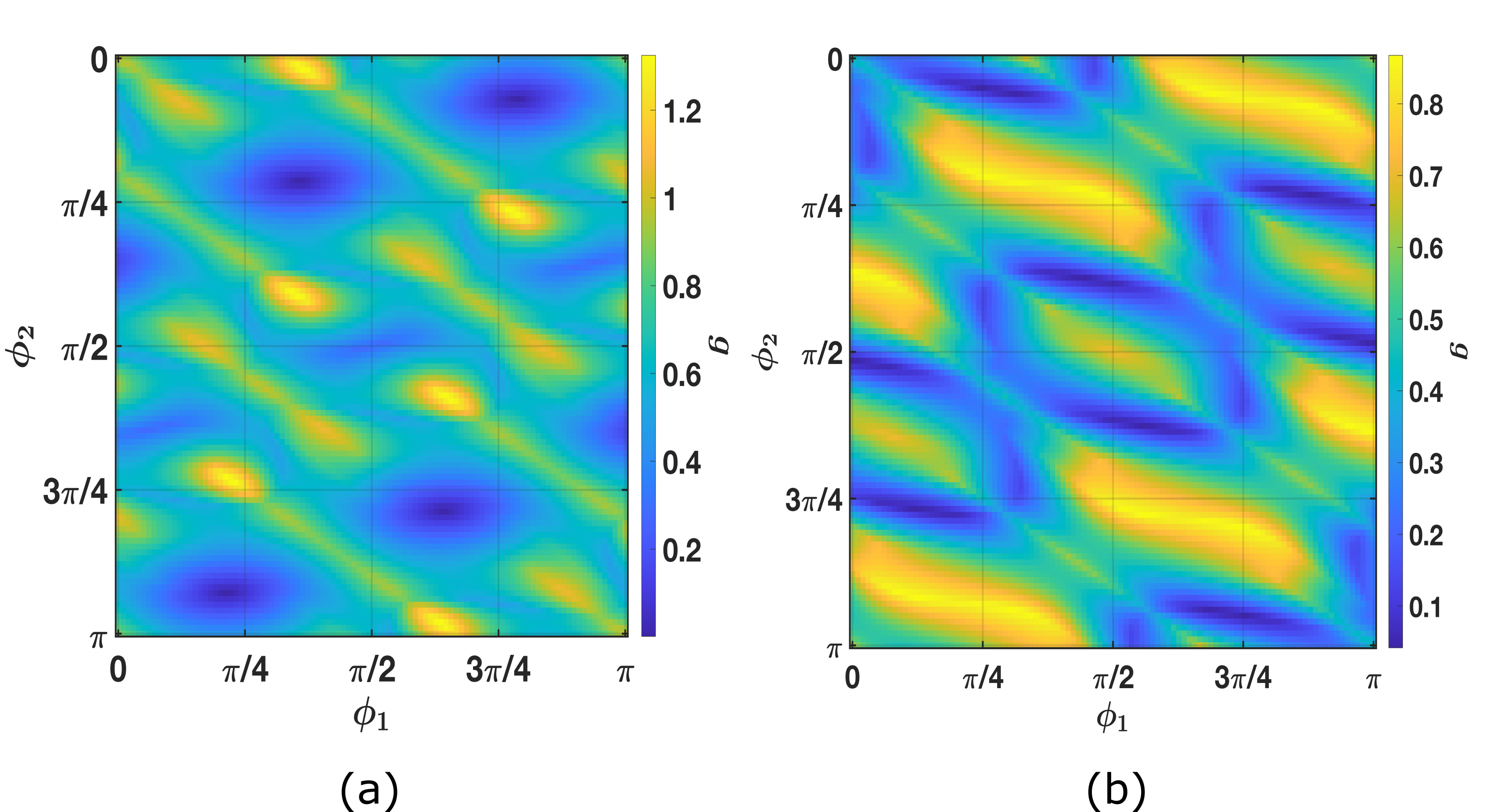}}
\caption{
Color-coded plot of
(a) Flatness ratio $g$ for the Valence band
(b) Flatness ratio $g$ for the Conduction band 
in $(\phi_1,\phi_2)$ plane for $v=w=t=1$.
}\label{fig:vb_cb_flat}
\end{figure}

We consider fluxes $\phi_1$ and $\phi_2$ impinging the octagon and square plaquettes as in Fig. \ref{fig:flux}. We plot the flatness ratios defined in the methods section of Valence and Conduction bands in the $(\phi_1,\phi_2)$ plane as shown in Fig. \ref{fig:vb_cb_flat}. Flatness ratios and their plots of other bands in the $(\phi_1,\phi_2)$ plane are presented in the Methods section. 
We observe that all energy bands display considerable flatness (according to the used metric defined in the methods section) hence can admit small Fermi-velocity of an initialized electronic wavepacket leading to non-trivial consequences. Recently such flat bands are being routinely investigated in twisted bilayer graphene \cite{lisi2021observation,marchenko2018extremely,morell2010flat} wherein Moire superlattices are formed at wavelength scale much bigger than the atomic separation in graphene. In these superlattices, hybridization of the energy bands from the two layers leads to flat bands when the twist angle is very low similar to what we see here.  Such bands have localized electronic density and can be the hot-bed for studying many correlated fermionic behavior as has been seen in a correlated Mott insulator under moderate occupancy \cite{po2018origin, yankowitz2019tuning, choi2019electronic, saito2020independent} or even a superconducting phase at low occupancy \cite{yankowitz2019tuning,codecido2019correlated,lu2019superconductors}. Even though we have described the electronic states only, the coupling of the Bloch states of the flat bands with spin can lead to many exotic spin-orbit interaction schemes like in Dzyaloshinskii-Moriya scheme \cite{cote2010orbital, PhysRevB.106.155402, PhysRevB.102.094404} and Ferromagnetic Mott state \cite{PhysRevLett.122.246402}. A subset of these phenomenon has been extended to transition-metal dichalcogenides \cite{pan2020band, PhysRevB.105.195428} too and there is no apparent reason why Poly-bi-[8]-annulenylene lattice cannot be the next fascinating test-bed. In fact it must be emphasized that unlike in graphene bilayer where mechanical twisting is necessary, herein we are able to generate flat bands under modest conditions using just a single chain with experimentally tailored magnetic flux profile.

\subsection*{Poly-bi-[8]-annulenylene in 2D}\label{2dsec}
\begin{figure}[h]
\centerline{
\includegraphics[width = 3.3 in]{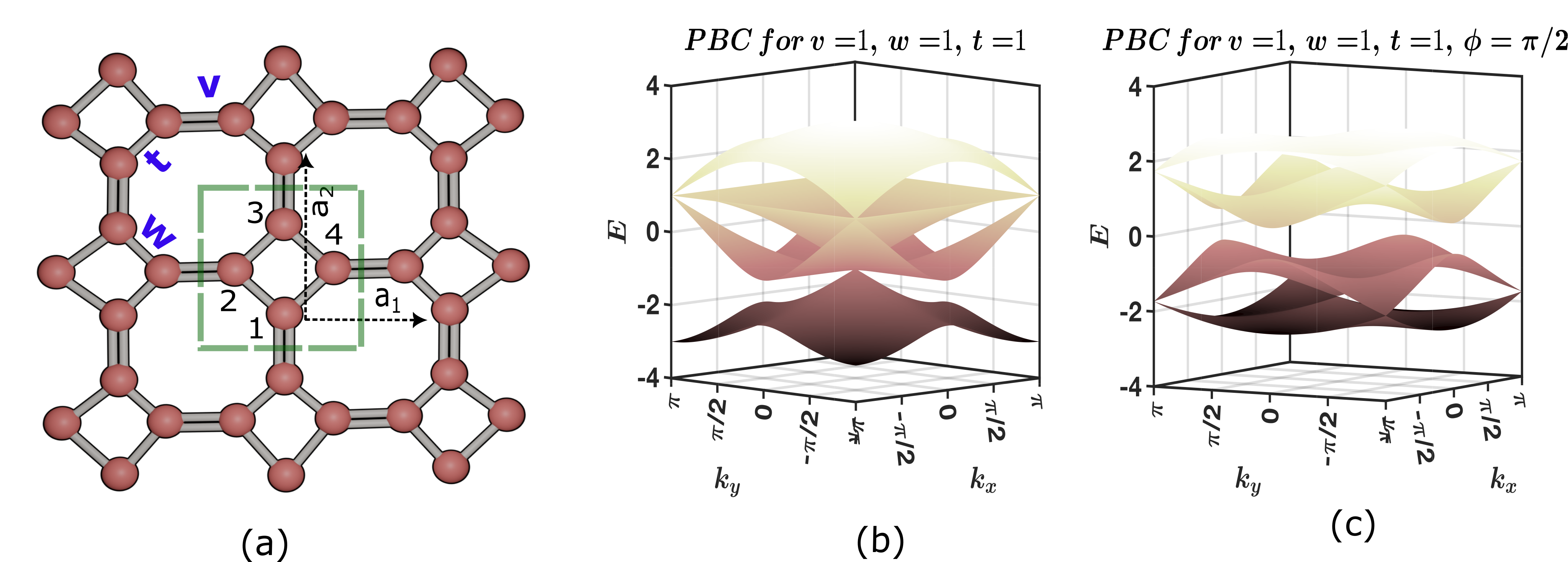}}
\caption{
(a) Square-Octagon lattice of Poly-bi-[8]-annulenylene in 2D with unit cell containing 4 sites boxed. 
(b) Bandstructure with $v=w=t=1$
(c) Bandstructure with $v=2$, $w=1$, $t=1$, $\phi=\pi/2$ as described by the new Hamiltonian Eq. \ref{eq:Ham_t_2d_sqoc_t_complex}
}\label{fig:sq_oct_lattice}
\end{figure}

The 2D lattice of Poly-bi-[8]-annulenylene has repeated COT units in both the x and y directions connected by squares. This lattice geometry has squares and octagons as its fundamental plaquettes. We investigate the lattice cleaved at $45^{0}$ as it reduces the lattice to a square with only four sites per unit cell, marked in red in Fig. \ref{fig:sq_oct_lattice}(a). The lattice vectors in Fig. \ref{fig:sq_oct_lattice}(a) are $\vec{a_1}=(1,0)$ and $\vec{a_2}=(0,1)$ with lattice constant taken to be unity. The Hamiltonian in the reciprocal space is given by,
\begin{equation}
H(k)=
\begin{pmatrix}
0 & w & ve^{-ik_y} & t\\
w & 0 & t & ve^{ik_x}\\
ve^{ik_y} & t & 0 & w\\
t & ve^{-ik_x} & w & 0\\
\end{pmatrix}
\label{tbeq}
\end{equation}
where $v$, $w$ and $t$ are nearest-neighbor hopping amplitudes as shown in Fig. \ref{fig:sq_oct_lattice}(a). Fig. \ref{fig:sq_oct_lattice}(b) shows the band structure of the model defined in Eq. \ref{tbeq} under PBC. Since the unit cell has four sites, there are four bands in the said figure. 
Our primary motivation is to explore the topological properties of the lattice and study what conditions lead to the emergence of zero-energy edge modes. To this end, we adopt the same technique as discussed in \cite{haldane} which involves raising the time-reversal invariance thereby culminating in band-opening at Time-reversal Invariant Momenta (TRIM) points($\Gamma$ \& $K$) \label{TRIM}\cite{Gilles}. This is akin to introducing a complex hopping parameter for nearest-neighboring interactions such that the resulting Hamiltonian has the following form,
\begin{equation}
\begin{pmatrix}
0 & we^{-i\phi} & ve^{-ik_y}e^{-i\phi} & te^{-i\phi}\\
w & 0 & te^{-i\phi} & ve^{ik_x}e^{i\phi}\\
ve^{ik_y}e^{i\phi} & te^{i\phi} & 0 & we^{-i\phi}\\
te^{i\phi} & ve^{-ik_x}e^{i\phi} & w & 0\\
\end{pmatrix} \label{eq:Ham_t_2d_sqoc_t_complex}
\end{equation}
where $e^{i\phi}$ is the complex hopping term introduced. In Fig. \ref{fig:sq_oct_lattice}(c) we show the resultant bandstructure of the model defined in Eq. \ref{eq:Ham_t_2d_sqoc_t_complex} wherein band-opening as discussed above is clearly evidentiated. 
\begin{figure}[h]
\centerline{
\includegraphics[width = 3.3 in]{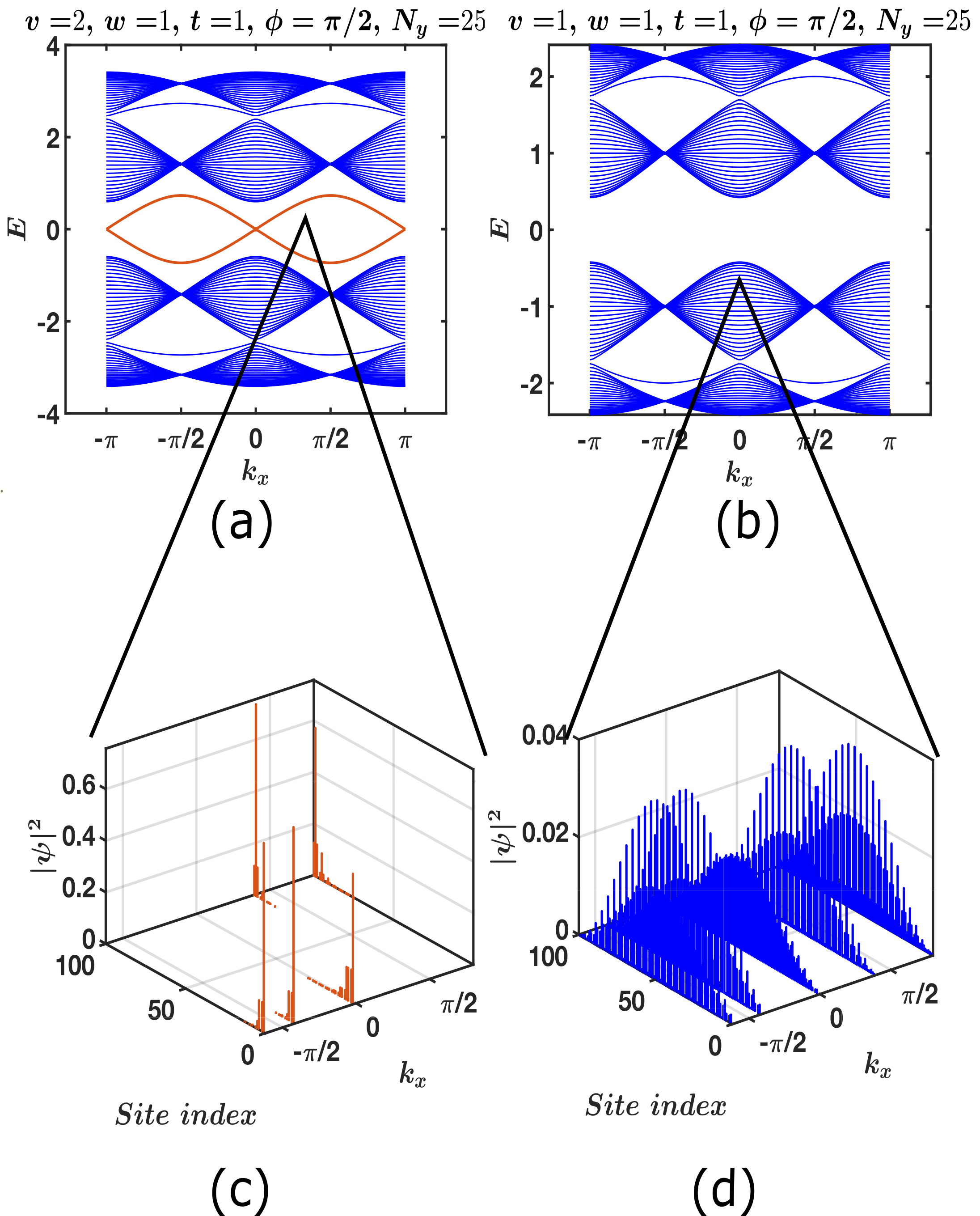}}
\caption{
(a) Spectrum of OBC along y- the direction and PBC in the x-direction. Marked in red: Edge modes
(b)Spectrum of OBC along y- the direction and PBC in the x-direction. Marked in red: Edge modes
(c) Electron density distribution of Edge Mode excitation
(d) Electron density distribution of a Bulk Mode.
($N_y$ denotes the number of finite unite cells taken along the y-direction)
}\label{fig:sq_tb_open}
\end{figure}

To envision the edge modes, it is sufficient to consider quasi-OBC i.e. periodicity in x-direction and aperiodicity in y-direction thus forming a cylindrical strip as explained in Methods. We show the appearance of zero energy eigenmodes corresponding to this type of edge in the eigenspectrum Fig. \ref{fig:sq_tb_open}(b), marked in red. We observe that such edge modes only occur in the topological phase for particular values of the parameters of the Hamiltonian after breaking the time-reversal symmetry of our system with complex coupling terms thereby indicating the formation of topological edge modes.

\section*{Conclusion}
We have studied the topology of COT and its associated polymeric structure: Poly-bi-[8]-annulenelene. We have presented ways to construct a number of topological excitations i.e. zero-energy edge modes both in strong as well as weak coupling regimes. By realizing the fact that the application of an external magnetic field affects the flatness of the bands, We have constructed systematic ways to tune the flatness of all the bands with uniform fluxes through every plaquette of the lattice geometry of 1D Poly-bi-[8]-annulenylene. The 2D extension of these polymers has Dirac points (similar to Graphene) and also supports topological phases upon lifting the time-reversal symmetry. The resilience of the incipient edge modes against noise has been discussed extensively (see supplementary). The discussed structures not only show promising conducting properties making them fundamental candidates for the Li/Na-ion batteries\cite{patent}-\cite{COTZhao2018SuperposedRC,Yin2020RecentPI} but also as we have shown in this article possess inherent topological characteristics and flatbands in the presence of an external magnetic field. Such insights are crucial to understanding its conducting properties and open the possibility of using these polymers as an alternative experimental ground to observe many flat-band-related phenomena as opposed to the previously used MOF/COF based platforms. 

Another interesting avenue which may benefit from a more careful investigation (to be undertaken shortly) is the prospect of simulating the physical effects studied in this article on a quantum hardware. With the recent advent of engineering lattices with superconducting qubits/cold atoms, COT and COT-based polymers could be engineered on table-top experiments and further exploited for their rich topological properties not only within a spinless fermionic model but can also be realized using a spin graph  phase as has been discussed explicitly in first Section of the Supplementary Information. In fact hybrid quantum simulation of materials and molecules and other physical systems have already begun to gain attention with interesting possibilities being explored \cite{sajjan2021quantum,sajjan2022quantum,sajjan2022magnetic,PhysRevResearch.5.013146,gupta2022hamiltonian, selvarajan2022dimensionality} including harnessing exotic correlation like entanglement \cite{li2019entanglement}. Taking a step further we show in this article that an engineered spin hamiltonian i.e. a Kitaev spin liquid(KSL) is capable of generating the same interaction as illustrated in Eq.\ref{eq:Ham_t_2d_sqoc_t_complex} after Jordan-Wigner transformation and majoranization. This opens up a lot of possibility for direct experimental simulation of the 2D-lattice on a superconducting hardware or even a cold-atom based quantum simulator\cite{ebadi2021quantum,bernien2017probing}. In short, we have just scratched the surface. The scope of possibilities to develop Poly-bi-[8]-annulenylene (in both 1D and 2D) as the next prospective candidate for beneficial applications as well as for procuring fundamental theoretical insight is practically endless. The authors hope that the findings in this article will duly bring into limelight other members of the tri-co-ordinated graphed lattices to unravel the unforeseen and untapped  chemistry these candidates are capable of displaying.

\acknowledgments
We thank Prof. Xingshan Cui from the Department of Mathematics, Purdue University for many useful discussions. The authors would like to acknowledge the financial support from the Quantum Science Center, a National Quantum Information Science Research Center of the U.S. Department of Energy (DOE).
\input{manuscript.bbl}

\newpage
\section*{Methods}
\subsection*{Calculation of Winding Number for COT}\label{sec:meth_wind}
The winding number of the COT envisioned as an SSH chain is a topological invariant  that characterizes its topological phase in 1D. This topological invariant described by the number of times the winding vector $\vec{d_k}$ of the SSH hamiltonian winds around the origin as shown in Figs.\ref{fig:ssh_main}(d)-(f) is given by\cite{Asboth},
\begin{equation}
    \gamma=\frac{1}{2\pi}\int_{-\pi}^{\pi} (\vec{d_k}\times\frac{d\vec{d_k}}{dk}) \,dk =
    \begin{cases}
      1, & |v/w|<1 \\
      0, & |v/w|>1
    \end{cases}
  \end{equation}
\subsection*{Pierls Substitution \& Flatness Ratios}\label{meth_flatness}
The pierls substitution terms in the hopping parameters of the hamiltonian and the flatness ratios computed to measure the flatness of the bloch bands are discussed in this section. Considering uniform fluxes $\phi_1$ and $\phi_2$ through the octagon and square plaquettes as shown in Fig. \ref{fig:flux} the resulting modification in the hopping terms along the octagon plaquette,
\begin{align*}
    e^{i\phi}[-v(c_{r,4}^\dagger c_{r,7}),-w(c_{r,8}^\dagger c_{r,7}),-v(c_{r,1}^\dagger c_{r,8}),-t(c_{r,2}^\dagger c_{r,1}),\\
    -v(c_{r,5}^\dagger c_{r,2}),-w(c_{r,6}^\dagger c_{r,5}),-v(c_{r,3}^\dagger c_{r,6}),-t(c_{r,4}^\dagger c_{r,3})]
\end{align*}\label{eq:oct_flux}
and along the square plaquette,
\begin{align*}
    e^{i\phi}[-w(c_{r,4}^\dagger c_{r,1}),-t(c_{r,3}^\dagger c_{r,4}),-w(c_{r,2}^\dagger c_{r,3}),-t(c_{r,1}^\dagger c_{r,2})]
\end{align*}\label{eq:sq_flux}
and their hermitian-conjugate(h.c) parts. \newline
The Flatness ratio $g=E_{bw}/\Delta$\cite{kai,titus,tang}, where $E_{bw}$ is the bandwidth and $\Delta$ is the bandgap is defined for the Valence band(VB) and Conduction Bands(CB). For all other bands, We consider the ratio $h=\langle V_F\rangle_K^{-(+)}/\langle V_F\rangle_K^{VB(CB)}$ where $\langle V_F\rangle_K$ denotes fermi-velocity averaged over the entire BZ and $-(+)$ denote negative(positive) Energy Bloch bands of the concerned system. The corresponding color-coded plots for the other bands are as follows.
\begin{figure}[t]
\centerline{
\includegraphics[width = 3 in]{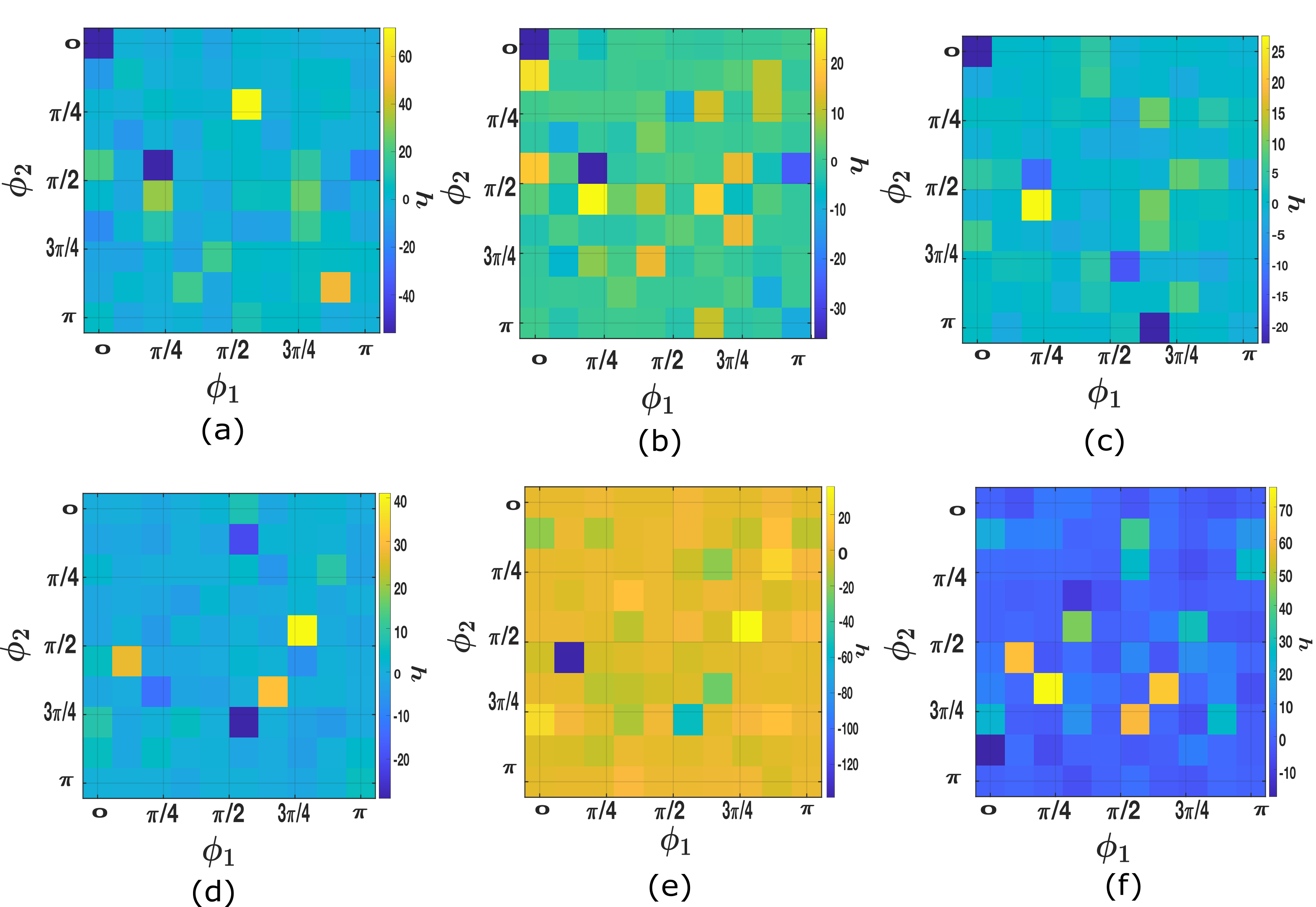}}
\caption{
Color-coded plot of
(a)-(c) Flatness ratio $h$ for the negative energy bands
(d)-(f) Flatness ratio $h$ for the positive energy bands
in $(\phi_1,\phi_2)$ plane for $v=w=t=1$.
}\label{fig:other_flat}
\end{figure}

\subsection*{Tight-Binding: Finite Size Calculations}
\begin{figure}[t]
\centerline{
\includegraphics[width = 1.5 in]{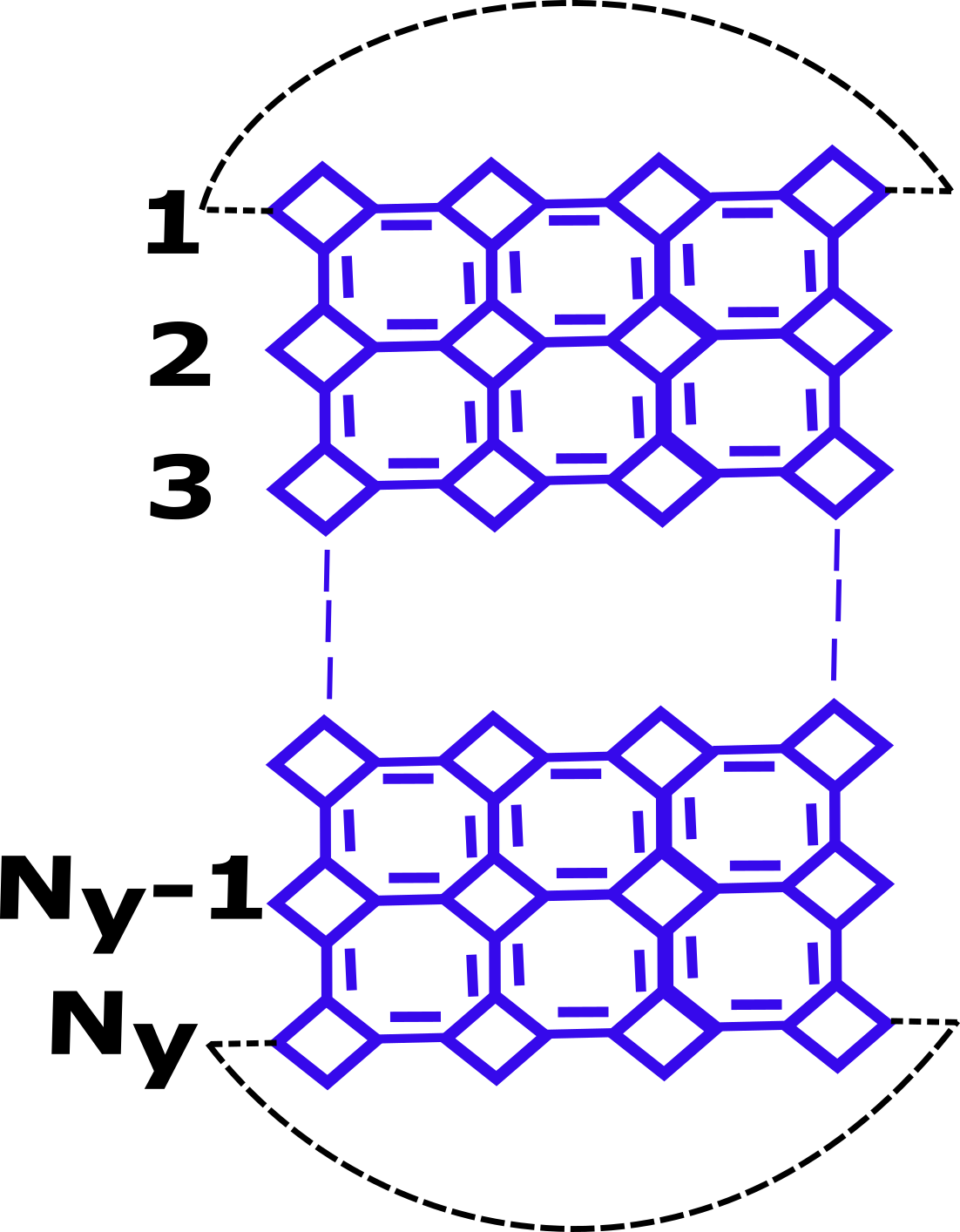}}
\caption{
Schematic of a cylindrical strip with PBC along x and OBC along y-directions. $N_y$ denotes the finite number of unit cells considered along the y-direction
}\label{fig:schematic}
\end{figure}

To study finite-size effects we consider quasi-OBC i.e. open boundary condition along y keeping the periodicity intact along x as shown in Fig. \ref{fig:schematic}. This forms a cylindrical strip with a finite number of cells in the y-direction alone thus forming zig-zag edges as shown.
For instance, for $N_y=4$ unit cells, the quasi-OBC hamitonian of PO[8]A in 2D has the form,
\begin{widetext}
\begin{align*}
\begin{pmatrix}
0 & we^{-i\phi} & 0 & te^{-i\phi} & 0 & 0 & 0 & 0\\
we^{i\phi} & 0 & te^{-i\phi} & ve^{ik_x}e^{-i\phi} & 0 & 0 & 0 & 0\\
0 & te^{i\phi} & 0 & we^{-i\phi} & ve^{-i\phi} & 0 & 0 & 0\\
te^{i\phi} & ve^{-ik_x}e^{i\phi} & 0 & we^{i\phi} & 0 & 0 & 0 & 0\\
0 & 0 & ve^{i\phi} & 0 & 0 & we^{-i\phi}  & 0 & te^{-i\phi}\\
0 & 0 & 0 & 0 & we^{i\phi}  & 0 & te^{-i\phi} & ve^{ik_x}e^{-i\phi}\\
0 & 0 & 0 & 0 & 0 & te^{i\phi}  & 0 & we^{-i\phi}\\
0 & 0 & 0 & 0 & te^{i\phi}  & ve^{i\phi}e^{-ik_x}  & we^{i\phi} & 0\\
\end{pmatrix}
\end{align*}
\end{widetext}

\section*{Supplementary Material}
\section*{Kitaev Spin Liquid(KSL) on 2D Poly-bi-[8]-annulenylene}\label{sec:appA}
In this section, we shall discuss a protocol to physically realize an effective Hamiltonian in Eq. \ref{tbeq} with a Kitaev spin liquid phase(KSL)\cite{kitaev2005,sq_octbaskaran,sq_oct_kells} which affords an interacting spin-graph with tunable interactions. We shall explicate analytically as well as numerically how the hopping parameters of the TB Hamiltonian can in turn be controlled by the parameters of the spin-spin interactions of the model. The Hamiltonian is given by,
\begin{equation}
    H_{kitaev}=-J_x\sum_{(i,j)\in x}\sigma_i^x\sigma_j^x-J_y\sum_{(i,j)\in y}\sigma_i^y\sigma_j^y-J_z\sum_{(i,j)\in z}\sigma_i^z\sigma_j^z
\end{equation}
The above hamiltonian in 2D lattice geometry of Poly-bi-[8]-annulenylene is given by\cite{kitaev2005}

\begin{align}
    H_{kitaev}=&-J_x\sum_r(\sigma_{r-a_1,4}^x\sigma_{r,2}^x+\sigma_{r,1}^x\sigma_{r-a_2,3}^x)\nonumber\\
    & -J_y\sum_r(\sigma_{r,1}^y\sigma_{r,2}^y+\sigma_{r,3}^y\sigma_{r,4}^y)\nonumber\\
    &-J_z\sum_r(\sigma_{r,2}^z\sigma_{r,3}^z+\sigma_{r,1}^z\sigma_{r,4}^z)\label{eq:jw_sq_oct}
\end{align}

\begin{figure}[h]
\centerline{
\includegraphics[width = 3 in]{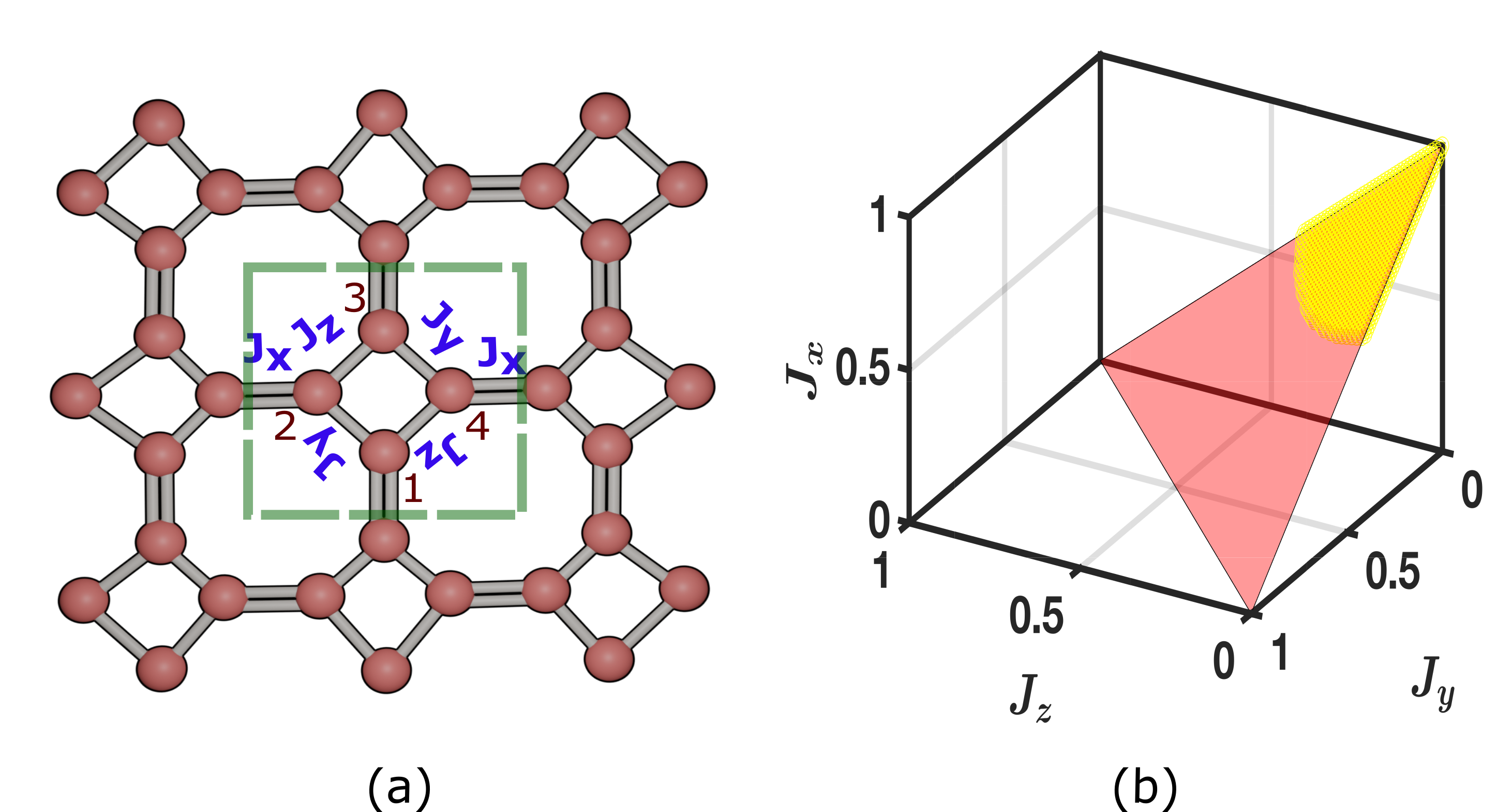}}
\caption{
(a) Spin graph of KSL on square octagon lattice with bonds: $(J_x, J_y, J_z)$
(b) Phase diagram of KSL on square octagon lattice on a fixed triangle $J_x+J_y+J_z=1$. The two insulator phases are shown in yellow and red with the yellow region hosting the Majorana edge modes.
}\label{fig:phase}
\end{figure}

Where $r,i$ denotes the position with $i=1,2,3,4$ covering all sites in the unit cell and $(J_x, J_y, J_z)$ are the bonds as shown in Fig. \ref{fig:phase}(a). Under Jordan-Wigner Transformation and majoranization following the Kitaev's protocol \cite{kitaev2005}, We have the Hamiltonian in k-space as,
\begin{equation}
    H(k)=
\begin{pmatrix}
0 & -i\frac{J_y}{4} & -i\frac{J_x}{4}e^{-ik_y} &  -i\frac{J_z}{4}\\
i\frac{J_y}{4} & 0 & -i\frac{J_z}{4} & i\frac{J_x}{4}e^{ik_x}\\
i\frac{J_x}{4}e^{ik_y} & i\frac{J_z}{4} & 0 & -i\frac{J_y}{4}\\
i\frac{J_z}{4} & -i\frac{J_x}{4}e^{-ik_x} & i\frac{J_y}{4} & 0\\
\end{pmatrix}
\end{equation}
The Majorana band structure of the above Hamiltonian has two gapped phases as shown in red and yellow in the phase diagram (see Fig. \ref{fig:phase}). For the topological phase (i.e. the yellow region in Fig. \ref{fig:phase}, The eigenstate density of the zero energy bands peaks at the edges of the chain and behaves as a Majorana edge state similar to that as shown in Fig. \ref{fig:sq_tb_open}(d). These two phases are indeed distinguished by a $\mathbb{Z}_2$ invariant\cite{yamada}.
\section*{Edge state resillience}\label{edgeresill}
In this section, We study the resilience of zero energy modes against the effects of the disorder\cite{Langnoise}. The disorder is introduced in the control parameters (i.e. the hopping amplitudes) as follows,
\begin{equation}
    T_{ij}\rightarrow T_{ij}+\delta T\,\epsilon
\end{equation}\label{eq:disorder}
Where $T_{ij}$ is a hopping amplitude in a clean Hamiltonian between sites $i$ and $j$. $\delta T$ is the strength of the disorder and $\epsilon$ is a random number sampled between $-1$ and $1$. We study two cases, 1) Resilience against a fixed $\delta T$ 2) Resilience against varying strengths of disorder $\delta T$. 
\subsection*{Poly-bi-[8]-annulenylene(PO[8]A) in 1D}\label{edgeresill_1d}
\begin{figure*}[h]
\centerline{
\includegraphics[width=13cm, height=10.4cm]{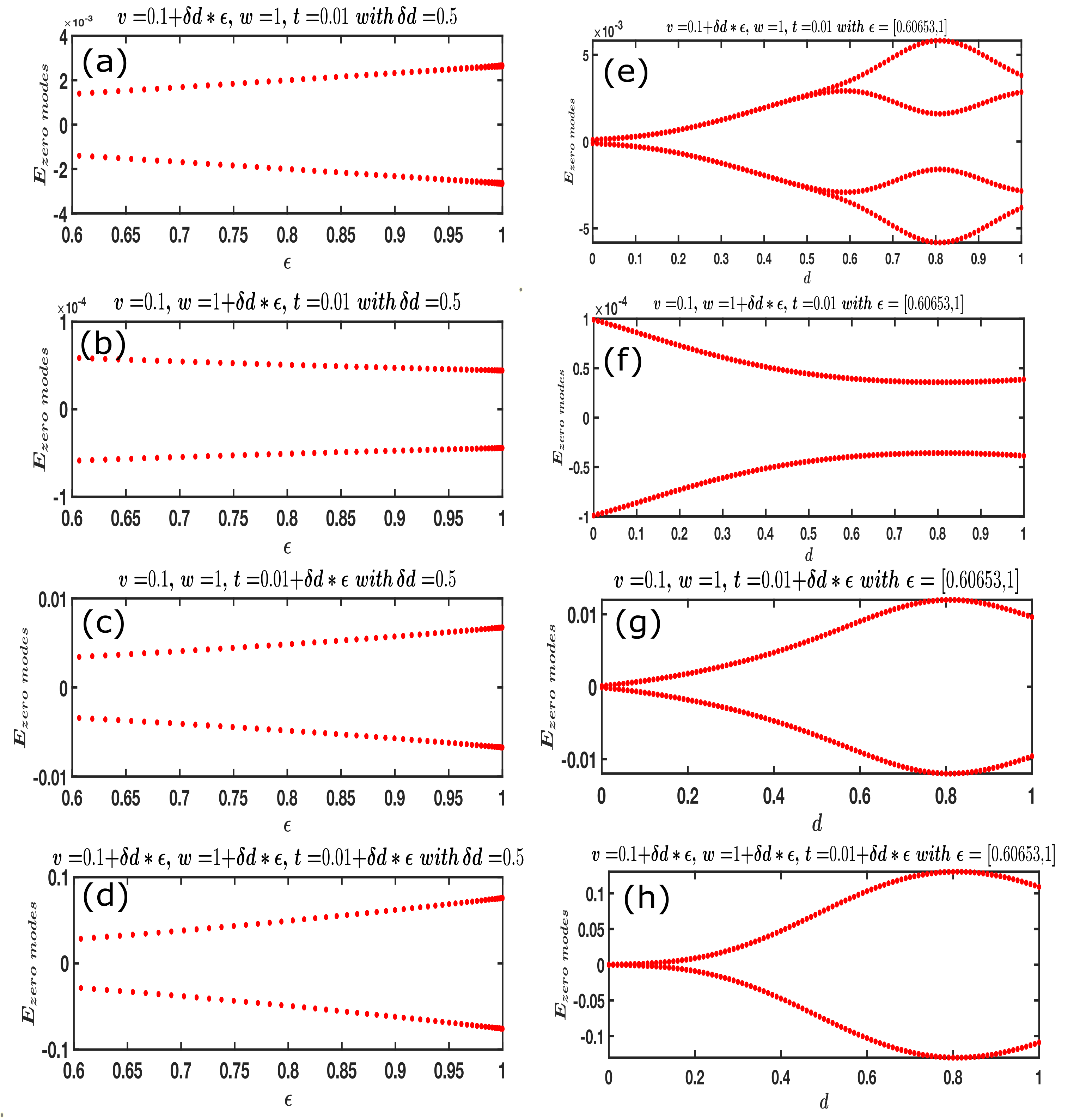}}
\caption{
(a)-(d) Resilience of zero modes with fixed Gaussian disorder strength
(e)-(h) Resilience of zero modes with varying strength of Gaussian disorder strength
}\label{fig:edgeress_1d}
\end{figure*}

\textit{Fixed strength:} In this case, We study the resilience of the zero energy modes with the disorder type of fixed disorder strength but a varying random number in the individual parameters of the Hamiltonian (Eq. \ref{eq:poly_model}) $v$, $w$ and $t$ respectively, analyzing the effects on both one parameter at a time Figs. \ref{fig:edgeress_1d}(a)-(d) and all the three simultaneously as in Fig. \ref{fig:edgeress_1d}(e).
\textit{Varying strength:} In this case, The disorder strength is varied along with a sampled random number in the individual parameters of the Hamiltonian (Eq. \ref{eq:poly_model}). Figures \ref{fig:edgeress_1d}(e)-(h) show the propagation of the zero energy modes under the above-mentioned disorder type.  

\subsection*{Poly-bi-[8]-annulenylene(PO[8]A) in 2D}\label{edgeresill_2d}
\begin{figure*}[h]
\centerline{
\includegraphics[width=13cm, height=10.4cm]{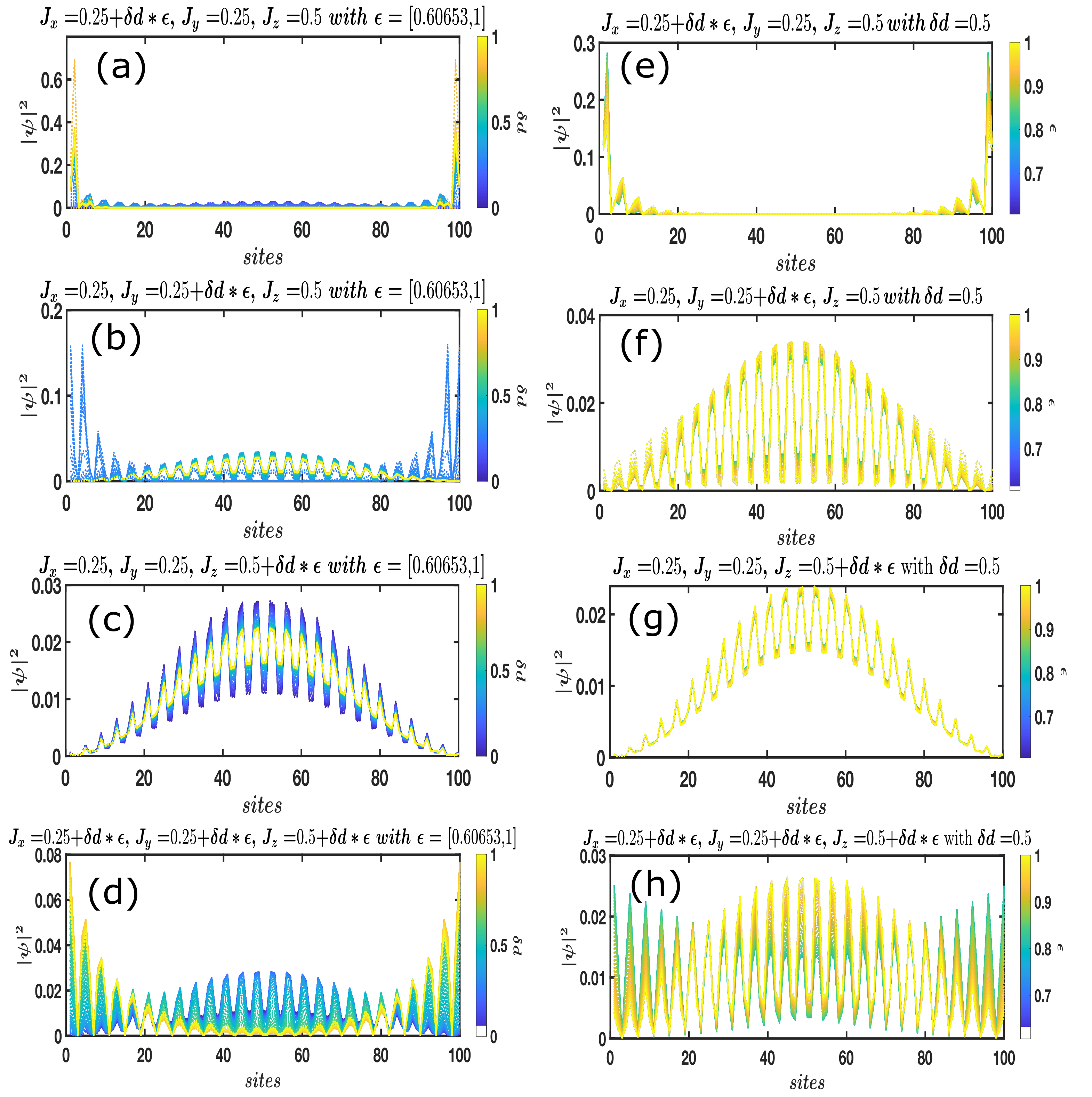}}
\caption{
(a)-(d) Resilience of edge state electron density with varying strength of Gaussian disorder
(e)-(h) Resilience of edge state electron density with fixed Gaussian disorder strength
}\label{fig:coher2}
\end{figure*}

For 2D lattice, The resilience of edge state electron density has been reported in Figs. \ref{fig:coher2}(a)-(d) for a fixed disorder strength and Figs. \ref{fig:coher2}(e)-(h) for varying disorder strength in different parameters of the Hamiltonian given by Eq. \ref{eq:jw_sq_oct}
\end{document}

%% file: manuscript.bbl
%